# Experimental realization of topologically-protected all-optical logic gates based on silicon photonic crystal slabs

*Furong Zhang[1], Lu He[1], Huizhen Zhang[1], Ling-Jun Kong[1], Xingsheng Xu[2], and Xiangdong Zhang[1]*

[1]Key Laboratory of advanced optoelectronic quantum architecture and measurements of Ministry of Education, Beijing Key Laboratory of Nanophotonics; Ultrafine Optoelectronic Systems, School of Physics, Beijing Institute of Technology, 100081 Beijing, China.

[2]State Key Laboratory of Integrated Optoelectronics, Institute of Semiconductors, Chinese Academy of Sciences, Beijing 100083, People's Republic of China.

**Abstract**

**Topological photonics has been developed for more than ten years. It has been proved that the combination of topology and photons is very beneficial to the design of robust optical devices against some disturbances. However, most of the work for robust optical logic devices stays at the theoretical level. There are very few topologically-protected logic devices fabricated in experiments. Here, we report the experimental fabrication of a series of topologically-protected all-optical logic gates. Seven topologically-protected all-optical logic gates (OR, XOR, NOT, XNOR, NAND, NOR, and AND) are fabricated on silicon photonic platforms, which show strong robustness even if some disorders exist. These robust logic devices are potentially applicable in future optical signal processing and computing.**

## 1. Introduction

All-optical logic gates are widely used in all-optical networks, signal processing and optical computing [1-3]. Traditionally, all-optical logic gates are made by nonlinear optical fibers [4, 5] and semiconductor optical amplifiers [6-8]. However, these traditional apparatuses are too large to be integrated, which limits their use. For realizing the future optical computer, as many all-optical logic gates as possible are expected to be integrated on chips. In order to obtain the compact footprint of the logic device, various methods and platforms are employed in recent decades, such as plasmon [9-11], photonic crystals (PhCs) [12-17], dielectric wire waveguide

[18-28], and so on. The plasmonic logic gates have an ultra-small footprint, but their optical loss is too large [9-11]. In contrast, the dielectric platform, especially the silicon on insulator (SOI) chip, usually possesses a negligible loss. Thus, all-optical logic gates based on the SOI chip have an overwhelming advantage in this respect. However, all these devices are vulnerable to perturbation. In complex environments, the performance and efficiency of logic gates are vastly decreased by environment disturbances. How to construct robust logic devices with low loss is important and necessary for practical application.

On the other hand, topological photonics have greatly received attention because of its robust optical transmission [29-42]. Even if some disorders exist, it also presents the feature of unidirectional propagating, due to the topological protection, such as the topological laser [43-45] and rainbow nano-photonic devices [46]. Combining topological photonics with the design of logic gates is expected to solve the above problem. In particular, the topological valley photonic crystals (TVPCs) show the feasibility of integrated optical logic chips due to their high coupling efficiency [31]. Although the theoretical scheme of the topologically-protected all-optical logic gates has been proposed [14], they have not been fabricated experimentally.

In this work, we experimentally fabricate a series of topologically-protected all-optical logic gates (including OR, XOR, NOT, XNOR, NAND, NOR, and AND) on the SOI chip. Especially, these gates show the properties of strong robustness against some disorders. It is believed that these topologically-protected logic devices possess wide application prospects in the fields of photonic integrated circuits.

## 2. Topologically-protected XOR and OR gates.

We consider the topologically-protected XOR and OR gates, which are fabricated by the Si PhC slabs. The scanning electron microscopy (SEM) image is shown in **Figure 1**(a). The lattice constant of the PhC slab is taken as $a$=450nm, and the thickness is 220nm. The unit cell is composed of two triangular air-holes in Si slabs. The PhC structures are symmetric about the plane with $z$=0 and surrounded by air at the top and bottom. Three kinds of cells are marked by orange (A), green (B), and blue (C), respectively, as shown in the right inset of Figure. 1(a). These triangle air-hole have different edge lengths, which are $d_1$=$d_2$ =0.5$a$ in B cells, $d_1$=0.3a (0.7$a$) and $d_2$ =0.7$a$ (0.3$a$) in A (C) cells. We construct the ABC-type and CBA-type supercells.

There are three layers of B lattices between A and C lattices, as shown in Figure 1(a). The band structures of the edge state for the supercells are plotted in Figure. 1(b), where the red (black) points represent the edge (bulk or leak) modes. Two edge states can be found in the bandgap. They respectively correspond to ABC- and CBA-type supercells [14, 47]. So, the topologically protected optical signals can propagate at the frequency of the bandgap. It should be noted that, as the number of layers of B cells increases, the topological band gap decreases slightly [47]. For this topological edge state, the optical propagating mode possesses a significant feature against some defects, like bending or deformation. That is because a non-trivial valley Chern number ($C_{K/K'}=\pm 1/2$) can be calculated in our proposed structure [29-33]. The detailed discussions of the band structures and topological properties can be found in S1 of Supporting Materials.

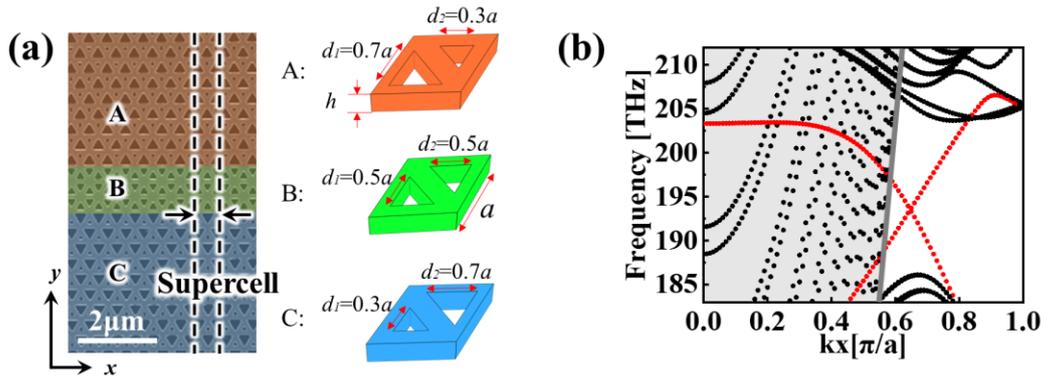

**Figure 1.** Topological valley photonic crystals. (a) The SEM image of Si valley photonic crystal slab and the unit cells. The thickness of the slab is $h$=220nm. (b) The band structure of the edge state for the ABC-type supercell.

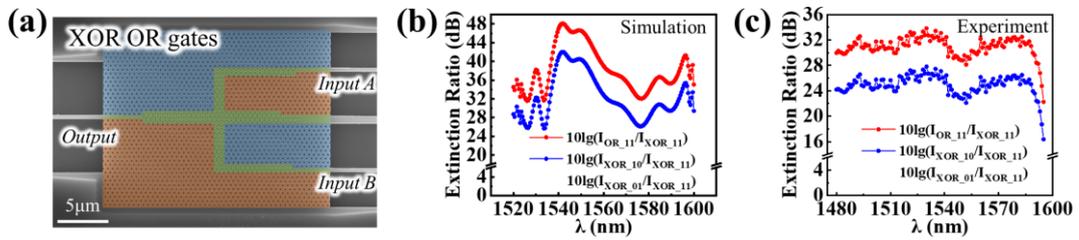

**Figure 2.** The XOR and OR gates. (a) The SEM image of the XOR and OR gates. (b) The simulation extinction ratio spectra between logic 1 (XOR_01, XOR_01, and OR_11) and logic 0 (XOR_11). (c) The corresponding experimental extinction ratio spectra.

Based on the robust edge states shown in Figure 1(b), various all-optical logic gates can be constructed. Here, we use the topologically protected ABC-type edge states to construct the XOR and OR gates. The schematic diagram is shown in **Figure 2**(a). The topological waveguide is designed as the "ψ" shape. For reducing the coupling loss, the additional coupling region is designed at input and output ports (See S2 of Supporting Materials for details). Based on the linear interference effect, the all-optical logic gates can be constructed. For instance, when Input A and Input B are both turned-on, two logic input signals are injected from the upper and lower right ports. They can propagate to the output port and interfere with each other. If the phase difference between Input A and B is adjusted as $\Delta\varphi=0$ ($\Delta\varphi=\pi$), the destructive (constructive) interference happens on the output port. In these two cases, the output port can be defined as XOR or OR port (named as XOR_11 and OR_11). If one of Input A and Input B is turned-on and the other is turned-off (XOR_01, XOR_10, OR_01, and OR_10 cases), the output signal is non-zero. If Input A and Input B are both turned-off (XOR_00 and OR_00), the intensity of the beam at the output port is zero.

From the above analysis, it is found that our designed device satisfies the logic operations of the XOR and OR gates. This can be demonstrated by numerical calculations. In general, the efficiency of the logic gate can be measured by calculating the extinction ratio (ER), which is defined as $10\log(I_1/I_0)$. Here, $I_1$ ($I_0$) represents the output intensity of the logic state 1 (0). The simulation results for the ER are shown in Figure 2(b). Red and blue lines represent the result of $10\lg(I_{OR\_11}/I_{XOR\_11})$ and $10\lg(I_{XOR\_10}/I_{XOR\_11})$, respectively. Due to the existence of spatial symmetry, the value of $10\lg(I_{XOR\_01}/I_{XOR\_11})$ is the same to that of $10\lg(I_{XOR\_10}/I_{XOR\_11})$. It can be seen that the maximum of the ERs is 48.07dB, which exhibits very good efficiency of logical function.

According to the designed structure, we experimentally fabricate the logic chip, the SEM image of the chip is also shown in Figure 2(a). Comparing it with the theoretical design, good consistency between them is confirmed. Now, we perform experiments to test the functions of XOR and OR gates. The detailed descriptions of the fabricated and measured methods are provided in Methods. The experiment results for ERs are plotted in Figure 2(c). The working bandwidth is ~100nm (ER>22dB) in the communication band (1480-1580nm). The maximum of the ERs is 33.83dB, which is much larger than the previous experimental results [18]. This means that a nice performance of XOR and OR gates has been demonstrated experimentally although there are some deviations from the simulation results. These deviations come from fabrication errors and the loss of optical measurement devices. In addition, there are some declining ERs closed to 1600nm, which is due to the measuring wavelength reaching the bandgap edge.

### 3. Cascaded XNOR, NAND, NOR, and AND gates.

For universal logic computing, the other complex logic gates are needed, which can be realized by cascading two or three logic units (XOR and OR gates). As shown in the SEM image of **Figure 3**(a), we design and fabricate the cascaded XNOR, NAND, and NOR gates. There are three input waveguides and one output waveguide, which are respectively named as Input A, Input B, Bias Light, and Output, connected to the topological logic devices. Now, we consider how these three cascaded logic gates work. For the XNOR gate, the phase difference between Input A and B is $\Delta\varphi=\pi$. Thus, the constructive interference makes logic signal propagate into the left logic unit from the right one. Meanwhile, the bias light is always turned-on, and its field amplitude is $\sqrt{2}/2$ times of that for Inputs A. When one of the input signals (Input A and Input B) is turned-on, the totally destructive interference happens at the output port with the output signal being 0. For other input states (the input signals are both turned-on or turned-off), the output amplitudes are nonzero, which corresponds to the logic output signal being 1. The experimental and simulation results for ERs are plotted in Figures 3(b) and 3(c), respectively. It turned out that the working bandwidth is ~100nm (ER>20dB) in communication band (1480-1580nm), and the maximal ER is up to 25.72dB in the experiment. Both experimental and theoretical results show good efficiency of the XNOR gate.

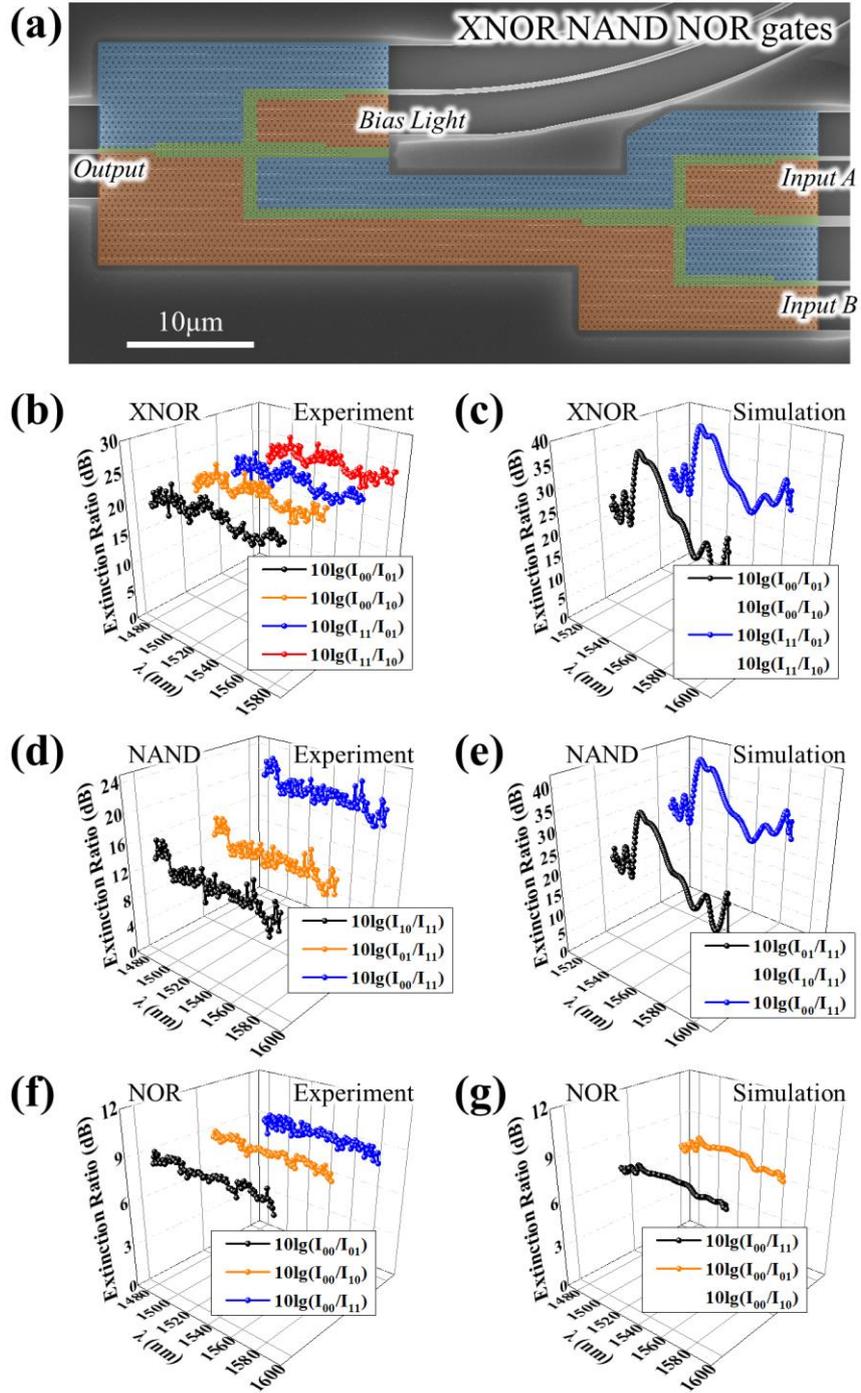

**Figure 3.** The cascaded XNOR, NAND, and NOR logic gates. (a) The SEM image of the XNOR, NAND, and NOR gates. The ERs of (b) and (c) for the XNOR gate, (d) and (e) for the NAND gate, (f) and (g) for the NOR gate. (b), (d), and (f) are the experiment results; (c), (e), and (g) are the simulation results.

For the NAND gate, it can be realized in the same way, except the amplitude at Bias Light is $\sqrt{2}$ times of that for Input A(B). When Inputs A and B are both turned-on, the completely destructive interference (zero output signal) appears. For the other three input states, the output amplitudes are nonzero. In this way, the logic function of the NAND gate can be realized. In Figure 3(d) and 3(e), we plot the experiment and simulation results of the ERs. The maximum of ERs is 22.88dB experimentally, and the working bandwidth is ~100nm (ER>10dB) in communication region. It means that the NAND gate is experimentally realized.

For the NOR gate, the amplitude at Bias Light is changed to $0.75\sqrt{2}$ times of that for Input A(B). When Input A and Input B are both turned-off, only the signal with Bias Light contributes to the output power. As to the other three situations, the destructive interferences happen at the output port, and the output powers are extensively reduced. The ratio between output powers of logic 0 and 1 is 9:1. Hence, the theoretical ER between logic 1 and 0 can be calculated as 9.54dB. The experiment results of the ER for the NOR gate (Figure 3(f)) have a good agreement with the corresponding simulation results (Figure 3(g)).

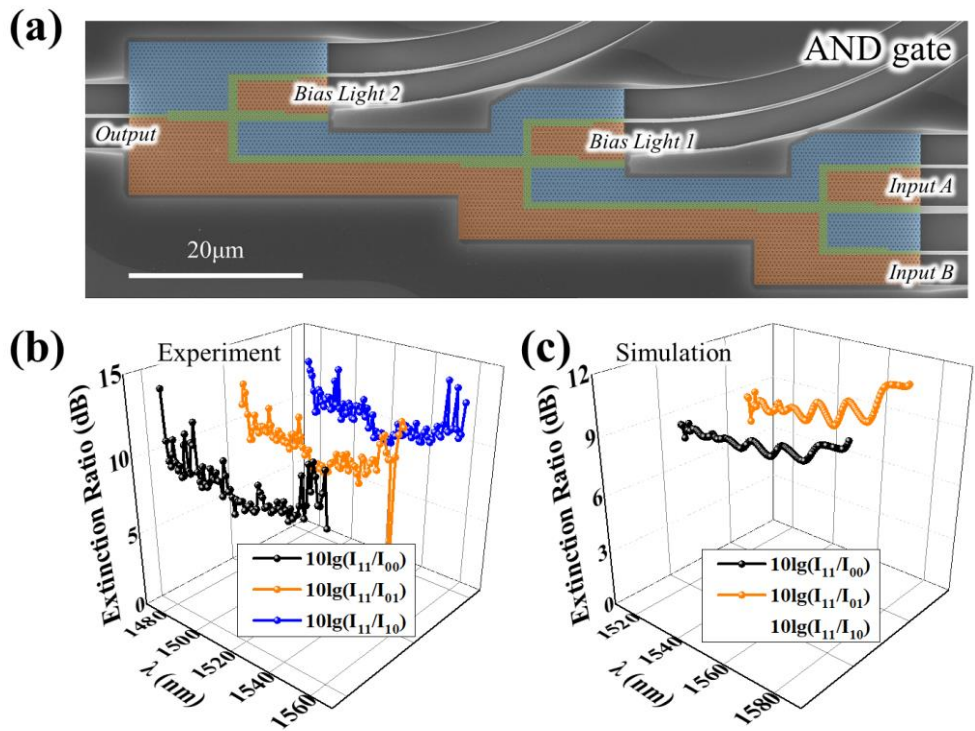

**Figure 4.** The AND gate. (a) The SEM image of the AND gate. The ERs of the AND gate: (b) the experiment results; (c) the simulation results.

The AND gate is constructed by the NAND gate and the NOT gate, as shown in the SEM

image of **Figure 4**(a). The amplitudes of the Bias Light 1 and Bias Light 2 are set as $\sqrt{2}$ and 0.75 times of that for Input A. The ratio of output powers for the AND gate is also 9:1 in theory. The simulation and experiment results of the ERs are shown in Figures. 4(b) and 4(c). In the communication band (1480-1560nm), the AND gate is basically realized. This means all major logic gates are experimentally fabricated around 1550nm. As shown in **Table 1**, we have listed the truth tables of all mentioned logic gates. The simulation results of above logic gates can be found in S3 of Supporting Materials. The important feature of these devices is robust against some disorders due to the topologically-protected edge state. In the following, we study the robust properties of these devices.

**Table 1.** The truth table of topologically-protected all-optical logic gates

| Inputs | | All major logic gates | | | | | | |
|---|---|---|---|---|---|---|---|---|
| Input A | Input B | OR (A+B) | XOR (A ⊕ B) | NOT ($\bar{A}$) | XNOR ($\overline{A \oplus B}$) $BL_1=\sqrt{2}/2$ | NAND ($\overline{AB}$) $BL_1=\sqrt{2}$ | NOR ($\overline{A+B}$) $BL_1=0.75\sqrt{2}$ | AND (AB) $BL_1=\sqrt{2}$ $BL_2=0.75$ |
| 0 | 0 | 0 | 0 | | 1 | 1 | 1 | 0 |
| 0 | 1 | 1 | 1 | 1 | 0 | 1 | 0 | 0 |
| 1 | 0 | 1 | 1 | | 0 | 1 | 0 | 0 |
| 1 | 1 | 1 | 0 | 0 | 1 | 0 | 0 | 1 |

**4. The robustness of the designed logic devices and its experimental demonstration.**

We provide the simulation results of a qualitative comparison of the robustness for topological and trivial logic gates [17] against disorders, as shown in **Figure 5**. In the simulation, a common disorder, waveguide bending, has been added in the devices. The trivial logic gates are still constructed using Si PhCs. They are made by the Si pillars with the lattice constant being 820nm, and the diameter of the Si rods being 460nm, which support the TM mode. The detailed description of the trivial PC-based logic gate can be found in S4 of Supporting Materials.

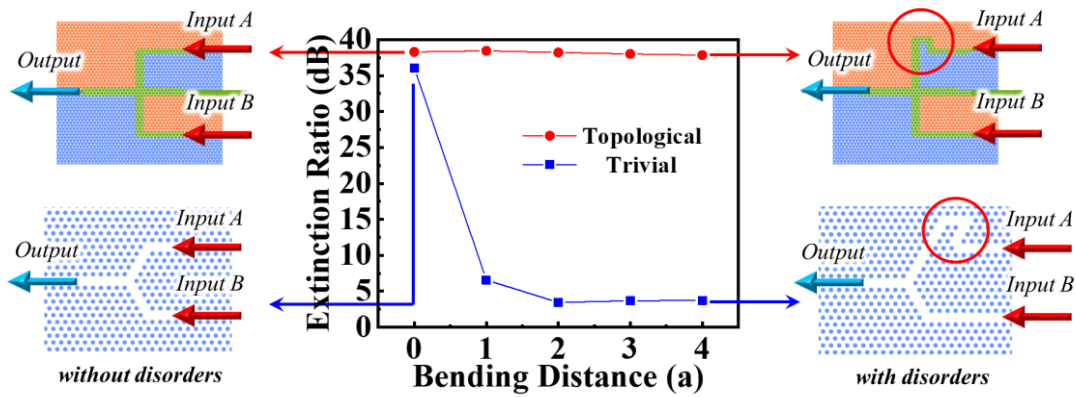

**Figure 5.** The quantitative comparison between the topological and trivial all-optical logic gate.

When no disorder exists, both topological and trivial logic gates can execute the logic function very well, as shown in the left part of Figure 5. And then, we introduce some defects in the upper input waveguide (the right part of Figure 5), the bending is gradually added in the upper input waveguide, which is marked by the red circles. The bending distance is set as the integral multiple of the lattice constant. We calculate the average ER in the working band (1520nm-1580nm). For the topological devices, the logic function can be maintained when the bending exists. In contrast, the performance is significantly reduced for trivial devices.

To further experimentally demonstrate the robustness of the fabricated logic devices, we fabricate the logic unit (the XOR and OR gates) with three defects, as shown in the SEM image of **Figure 6**(a). Three red circles mark the disorders, which are bulging, bending, and indentation [14]. For the certain incident optical signal with the constant intensity and phase, the device possesses a good protection against the disorder, because the topological protection can reduce the scattering loss. Such a device can also protect linear interference effects actually. For example, when an optical signal is propagating through a bending waveguide, an additional phase is generated at the waveguide bending. Thus, the interference effect becomes worse, because the phase difference is no longer zero. However, we have found some ways to deal with this problem. Once the structure is determined, the additional phase is a fixed value regardless of the shape and size of the waveguide bending. Therefore, we can do a pre-measurement to find out what value the fixed phase difference of the two input lights is. When two optical signals inputted into the logic chip, we can consider such a way to avoid the imperfect interference effect.

By using the above measurement methods, we obtain the experiment results of the ERs, as shown in Figure 6(b). Even though the perturbations exist, the logic function can also be executed. A large ER (more than 20dB) can be seen in the communication band (1480-1590nm). Comparing them with those in Figure 1(d), we find that the logical function has not been greatly affected with the introduction of disorders. The experimental results demonstrate that these devices have good performance though under the presence of significant disorders.

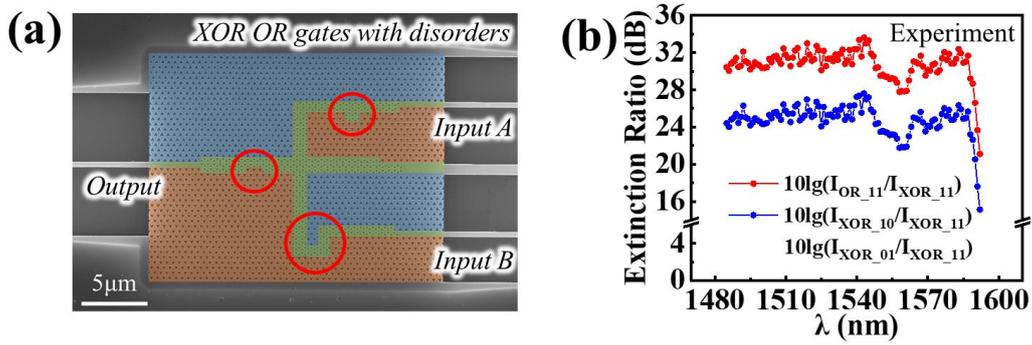

**Figure 6.** The XOR and OR gates with disorders. (a) The SEM image of the XOR and OR gates with three defects. (b) The experiment results of ERs for the XOR and OR gates.

As for the other type of defect, it may exist in the structure caused by complex environments, such as small particles and dust. Our simulation results clearly show a strong disorder tolerance for the logic gates. Detailed discussions are given in S5 of Supporting Materials. In addition, we would like to emphasize that the previous work [48] focuses on theoretical design of logic gates. It employs two-dimensional photonic crystals to design logic gates. This makes the experiment difficult to implement. Because in the communication band, two-dimensional photonic crystals are difficult to be fabricated experimentally. In contrast, the present work focuses on two-dimensional photonic crystal slabs with SOI chips, which enables the experiments to implement these logic devices with topology protection.

## 5. Conclusion

To conclude, we have designed and fabricated the topologically-protected all-optical logic gates based on the SOI chip. All major topologically-protected logic gates (OR, XOR, NOT, XNOR, NAND, NOR, and AND) with low loss and highly efficient performance have been demonstrated experimentally. The topologically-protected all-optical logic gates can work well though under the presence of significant disorders, which shows strong robustness against disorders. We believe that the topologically-protected all-optical logic gates have extensive application prospects in the future optical computing field.

## 6. Methods

**Sample fabrication.** Firstly, the photonic crystal slabs were fabricated using electron beam lithography, followed by dry etching. The substrate was a silicon-on-insulator wafer with a 220 nm-thick top Si layer. ZEP-520A e-beam resist was firstly spun-coated on the substrate for exposure, and resist patterns were formed after e-beam lithography and development. Then these resist patterns were transferred to the top Si layer using inductively coupled plasma etching in $SF_6$ and $CHF_3$ gases atmosphere, with ZEP520A used as an etching mask. The etching depth for logic gates and waveguides is 220nm.

In the second step, the $SiO_2$ under the photonic crystal slab was removed by wet etching process. We used a patterned photoresist (S1813) as a mask for the wet etching. S1813 photoresist was firstly spun onto the sample, and resist patterns were formed after ultraviolet lithography and development. Next, the patterned photoresist was hardbaked on a hotplate for 60 minutes at 115°C. Then the sample was rinsed in a buffered oxide etch (BOE), and the $SiO_2$ under photonic crystal slab was etched away with the patterned photoresist used as etching mask. Finally, the photoresist was stripped away with acetone, followed by $O_2$ plasma treatment to clean the residual photoresist thoroughly.

**Measurement method.** In our experiment, the incident light comes from a continuous wave laser (from 1480nm to 1630nm). In free space, incident light is divided into four paths. In order to control the polarization and intensity ratio of the lights, a series of wave plates were placed in the light path. Then, the light beams enter the chip through the fiber array. The output signal is detected by a high-speed optical power monitor. It is note that phase of beam in fibers is unstable due to the eigenvibration of fibers and temperature variation. A dynamic measured method is employed to test output signals of our logic gates. Here, in the measurement, we used three fiber phase modulators [49] with different modulation periods. More details can be found in S6 of Supporting Materials.

Furthermore, we also propose a scheme to reduce the phase instability by employing the optical phase lock loops (OPLLs) [18]. The detailed scheme is discussed in S7 of Supporting Materials.

## Acknowledgements

This work was supported by the National key R; D Program of China under Grant No. 2022YFA1404904, National Natural Science Foundation of China (12234004, 12004038 and 11904022).

## Conflict of Interest

The authors declare no conflict of interest.

## Author Contributions

F. Z. and L. H. contributed equally to this work. L. H. finished the theoretical scheme and designed the topologically-protected all-optical logic gates. F. Z. and L. H. finished the experiments with the help of L. K. and X. X.. L. H. analyzed experimental data. L. H. and X. D. Z. wrote the manuscript. X. D. Z. initiated and designed this research project.

## Data Availability Statement

The data that support the findings of this study are available from the corresponding author upon reasonable request.

## Keywords

Topological protection, logic gate, optical computing, photonic crystal slabs

# Supporting Materials for

# Experimental realization of topologically-protected all-optical logic gates based on silicon photonic crystal slabs

Furong Zhang[1],† Lu He[1],† Huizhen Zhang[1&], Ling-Jun Kong[1], Xingsheng Xu[2], and Xiangdong Zhang[1]*

[1]*Key Laboratory of advanced optoelectronic quantum architecture and measurements of Ministry of Education, Beijing Key Laboratory of Nanophotonics; Ultrafine Optoelectronic Systems, School of Physics, Beijing Institute of Technology, 100081 Beijing, China.*

[2]*State Key Laboratory of Integrated Optoelectronics, Institute of Semiconductors, Chinese Academy of Sciences, Beijing 100083, People's Republic of China.*

†These authors contributed equally to this work. *[&] Author to whom any correspondence should be addressed: zhangxd@bit.edu.cn; zhanghz96@126.com

## S1. The band structure of the topological valley photonic crystals.

In the main text, we consider the topological valley photonic crystals (TVPCs) to design the all-optical logic gates. Here, we show the details of the TVPCs. The lattice constant is $a$ = 450nm. The unit cells are composed of two triangular air-holes in Si slab, the thickness is $h$=220nm. The slab is surrounded by air at the top and bottom, and symmetric about the $z$=0 plane. In Fig. 1(a), three kinds of cells are presented and marked in orange (A), green (B), and blue (C), respectively. Their edge lengths of the triangle air-hole are different, which are $d_1$=$d_2$ =0.5$a$ in the B cell, and $d_1$=0.4$a$ (0.6$a$) and $d_2$ =0.6$a$ (0.4$a$) in the A (C) cell. The bulk band (the red points) of B cell is gapless at K(K') point due to the $C_6$ symmetry, as plotted in Fig. 1(b). As for the A and C cells, the edge length $d_1$ ($d_2$) is changed to 0.4$a$ or 0.6$a$ for opening the band gap, as shown by the blue band in Fig. 1(b).

Moreover, the magnetic-field distributions (Hz) can also be calculated. Based on the **k·p** model, the effective Hamiltonian around K/K' valley can be expressed as $H_{K/K'} = \pm(v_D \delta k_x \sigma_x + v_D \delta k_y \sigma_y) \pm m v_D^2 \sigma_z$, where $v_D$ is the group velocity, and δ**k**=**k**-**k**$_{K/K'}$ is the displacement from the wave vector **k** to K/K' valley in the reciprocal space, $\sigma_x$, $\sigma_y$, and $\sigma_z$ are the Pauli matrices, $m$ is effective mass. It is known that the valley Chern number is $C_{K/K'}$=±sgn($m$)=±1/2. It is clearly confirmed by the power flux directions in Fig. S1(c). To illustrate this effect, we calculate profiles of Berry curvature for the first band of A and C cells in momentum space, as shown in Fig. S1(d). It can be seen that the Berry curvatures with opposite signs appear at K(K') points, manifesting the different valley topologies for two structures.

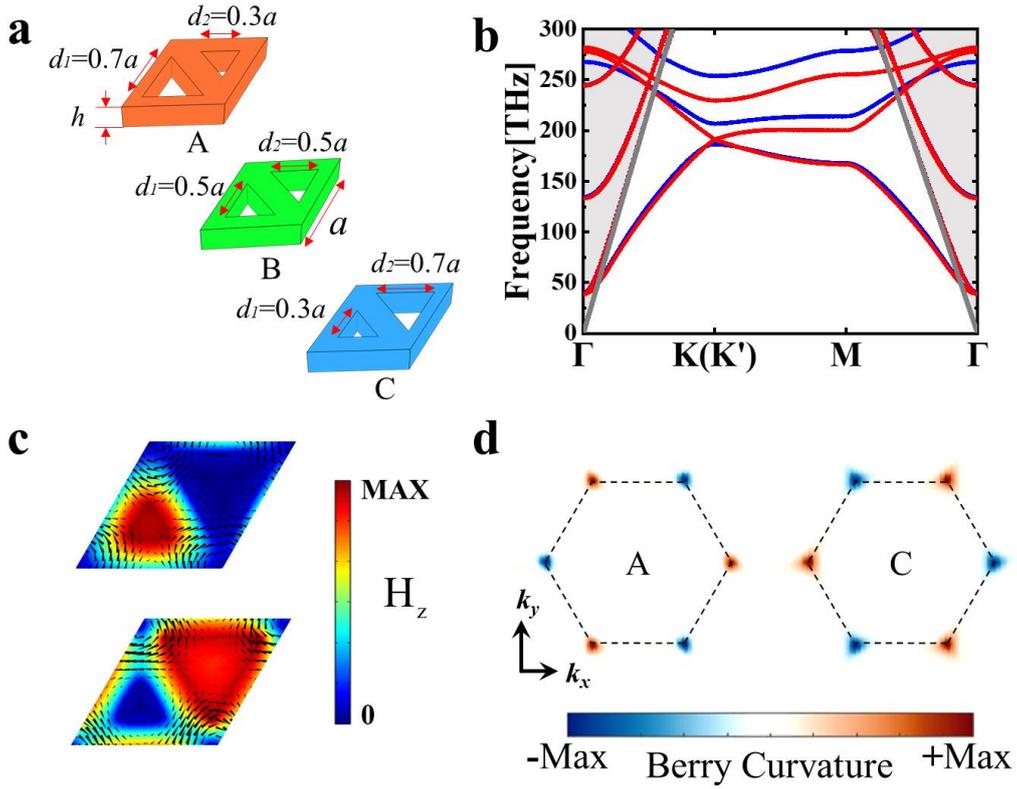

Fig. S1. The unit cells and their band structures. (a) The unit cells. (b) The band structure of (a). The red line corresponds to cell B, the blue line corresponds to cell A or C. (c) The magnetic-field distributions ($H_z$) of cell C at K point. Black arrows: power flux. (d) Berry curvature distribution of the first band of A and B. The black dashed line represents the first Brillouin zone.

**S2. The design of the coupling region from the wire waveguide to the topological channel.**

Here, we design the coupling region between the Si wire waveguide and the edge state of the topological photonic crystal, which is inspired by Ref. [31] of the main text. For the coupling region of the ABC and CBA topological edge states, different waveguide widths are required to maximize their coupling efficiencies. In Fig. S2(a), the ABC-type topological edge state employs the coupling waveguide of one lattice constant. In this way, the overlapping integral between the eigenmode of the topological edge state and the fundamental mode of the waveguide (TE0) can be maximized. As a result, coupling efficiency is maximized. In Fig. S2(b), we numerically calculate the coupling model from the waveguide to the ABC-type topological edge state. Results show that the coupling efficiency is close to 100%. As shown in Fig. S2(c), for the CBA-type edge states, we also designed the coupling region according to the principle of maximizing the overlapping integral between the eigenmode of the topological edge state and the fundamental mode of the waveguide (TE0). We find that coupling efficiency can be maximized by using waveguide coupling with two lattice lengths. In Fig. S2(d), we numerically calculate the coupling model from the waveguide to the CBA-type topological edge state. Results show that the coupling efficiency is also close to 100%.

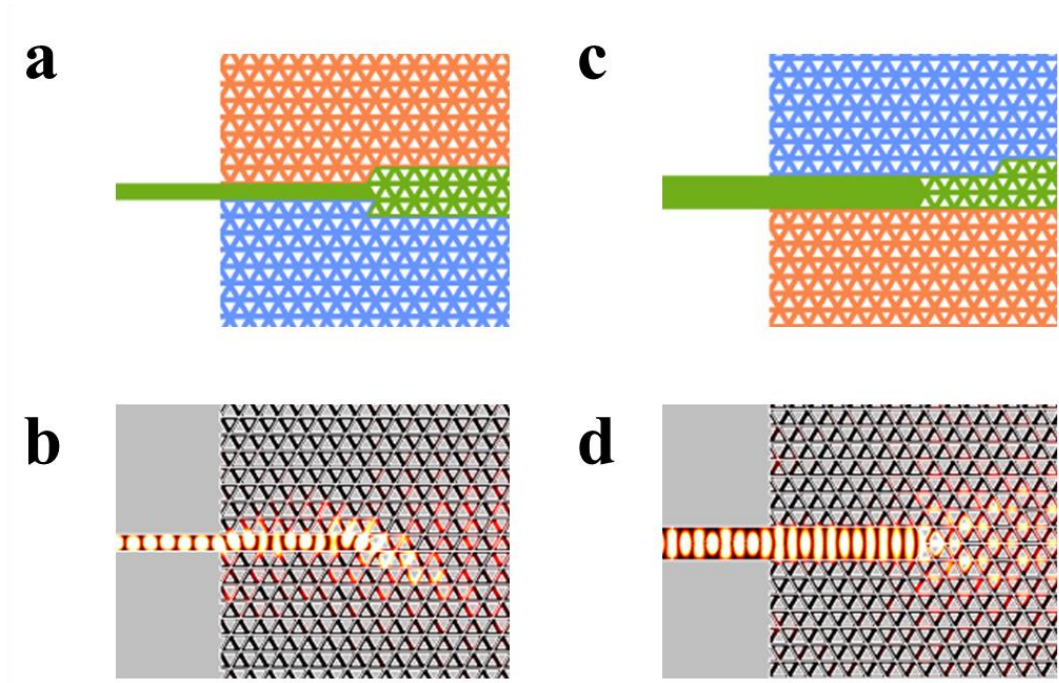

Fig. S2. The coupling region from the wire waveguide to the topological edge state. (a) The scheme of the coupling region of ABC-type edge state. (b) The simulation result of coupling for ABC-type edge state. (c) The scheme of the coupling region of CBA-type edge state. (b) The simulation result of coupling for CBA-type edge state.

### S3. The simulation results of the all-optical logic gates.

In this section, we calculate the simulation results of the all-optical logic gates, including XOR, OR, NAND, XNOR, NOR, and AND gates (Ref. [14] in the main text).

Based on the discussion in the main text, the XOR and OR gates can be realized by using the linear interference effect. When the phase difference of Input 1 and 2 are set as $\Delta\varphi=0$ ($\Delta\varphi=\pi$), the Output will be defined as XOR (OR) port. the destructive and constructive interference happens on the XOR and OR Outputs respectively, as shown in Figs. S3(a) and S3(b). When one of the Inputs is turned-on and the other one is turned-off, the output amplitude is non-zero at the Output port. This case (logic state 01 or 10) corresponds to Figs. S3 (c) and S3(d).

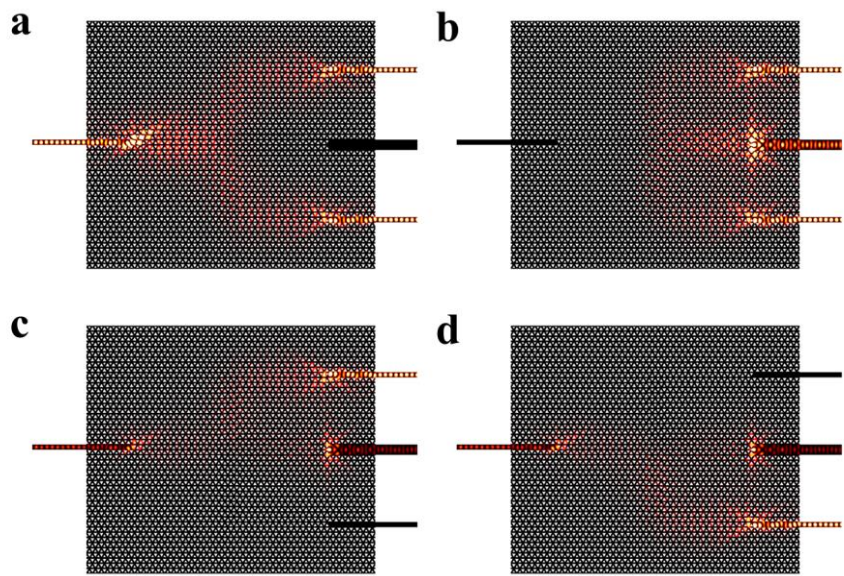

Fig. S3. The simulation results of the XOR and OR gates.

In Fig. S4-S5, we show the simulation results of XNOR, NAND, NOR, and AND gates. The logic input states 00, 01, 10, 11 correspond to (a), (b), (c), and (d), respectively.

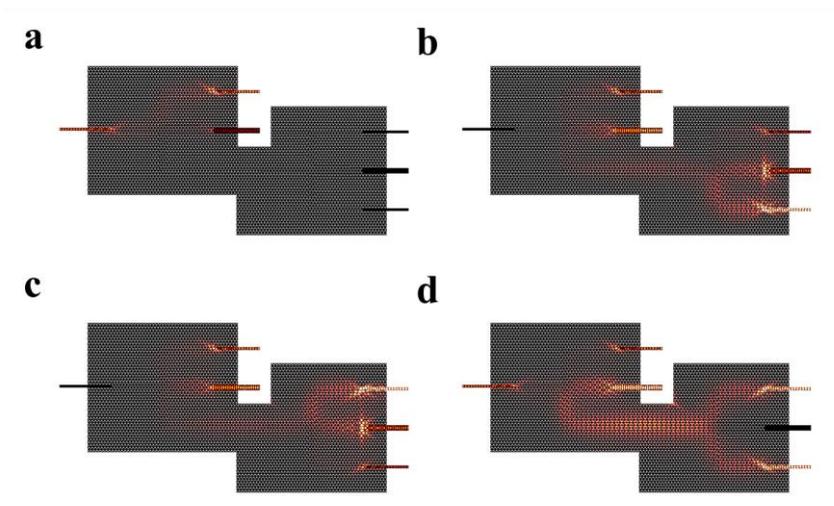

Fig. S4. The simulation results of the XNOR gate.

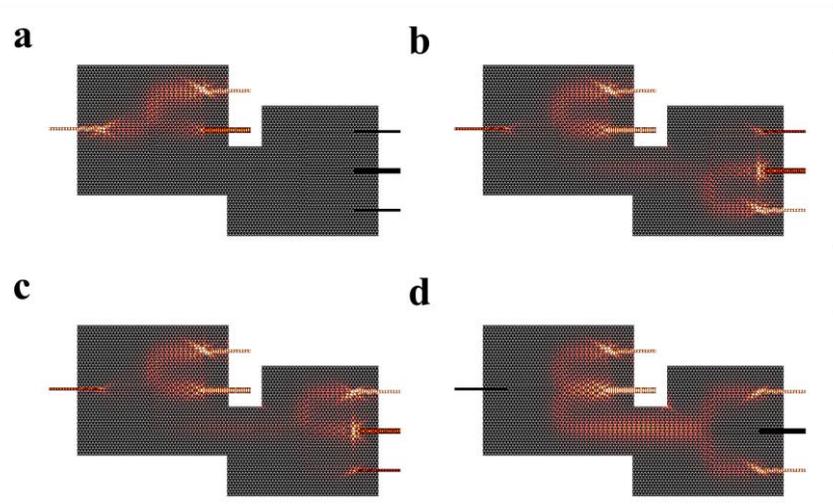

Fig. S5. The simulation results of the NAND gate.

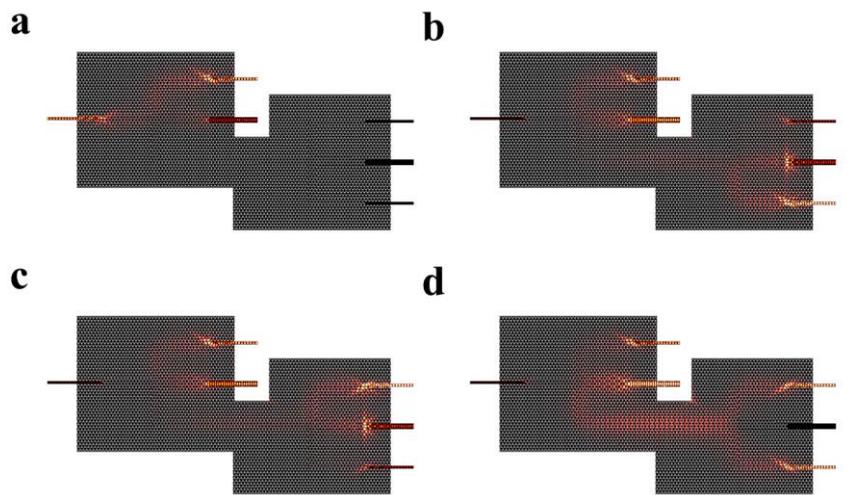

Fig. S6. The simulation results of the NOR gate.

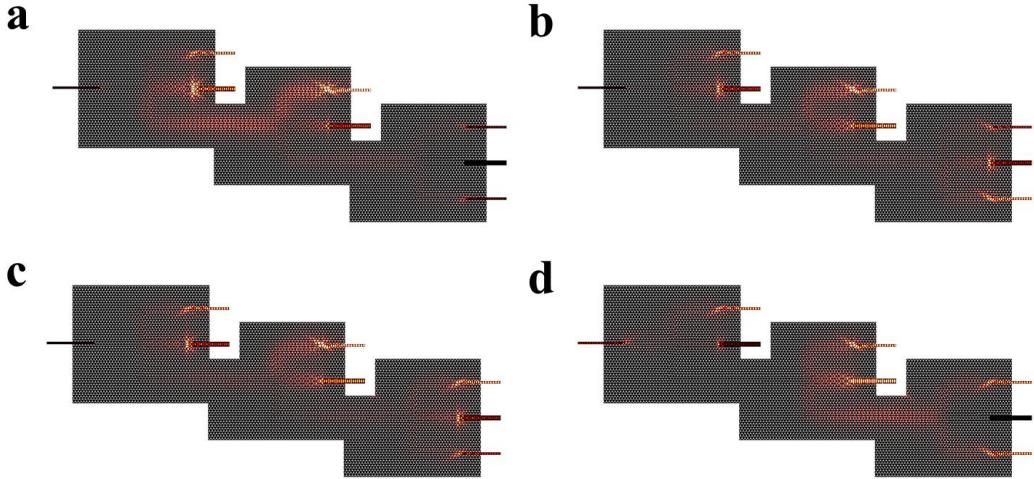

Fig. S7. The simulation results of the AND gate.

## S4. The simulation results of the topological and trivial all-optical logic gates with disorders.

We use the trivial all-optical logic gate as a control group in Ref. [17] of main text, as shown in Fig. S8(a). The lattice constant is a = 820nm. By the simulation calculating, Fig. S8(b) shows that a nice performance of ERs (marked as blue and cyan lines) for the trivial logic gate can be gotten without the defect. And then, we introduce some defects in two input waveguides, as shown in the right part of Figs. S8(d) and S8(f). However, the bending defect results in a bad performance of ERs (marked as red and pink lines), as shown in Fig. S8(b). These results further demonstrate our topological device can be widely used in the complex environment of the future optical computing, and the trivial device cannot.

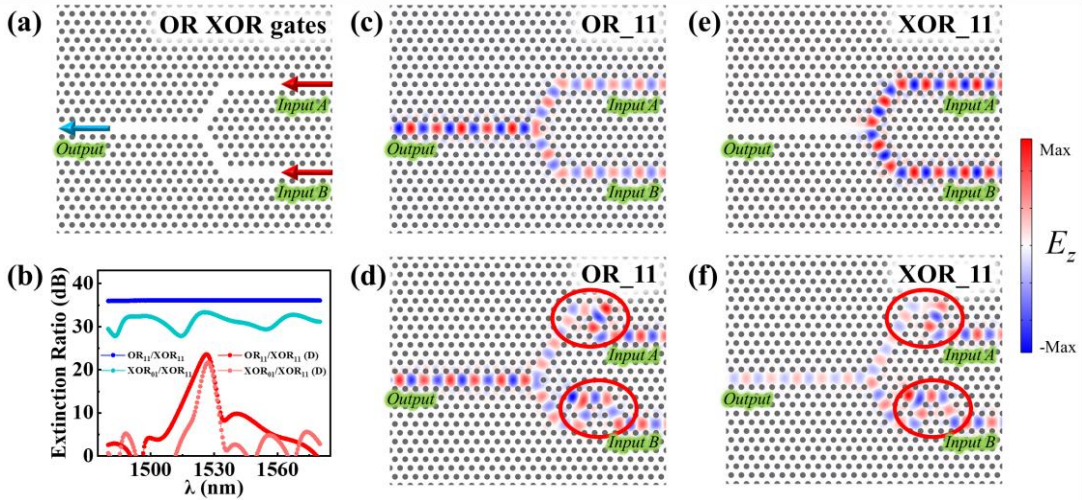

Fig. S8. (a) The trivial PCs logic gate without and with the defects. (d) The simulation results of ERs. (c) The calculated electric-field distributions ($E_z$) of logic states (c) OR_11, (d) OR_11 with disorders, (e) XOR_11, (f) XOR_11 with disorders.

## S5. The simulation results of the topological all-optical logic gates with air-hole missing disorders.

Here, we consider the air-hole missing disorder in these devices, as shown in Fig. S9a. Black points represent the missing air hole. Here, we remove 3 air holes in the topological edge state. The enlarged image clearly shows the positions of missing air holes. In order to characterize the efficiency of the logic gate, we calculate the extinction ratio (ER) of the simulation results, which is defined as $10\log(I_1/I_0)$. Here, $I_1$ ($I_0$) represents the output intensity of the logic state 1 (0). The simulation results for the ER are shown in Fig. S9(b). Red and blue lines represent the result of $10\lg(I_{OR\_11}/I_{XOR\_11})$ and $10\lg(I_{XOR\_10}/I_{XOR\_11})$, respectively. It can be seen that the maximum of the ERs is up to 50dB, which exhibits very good efficiency of logical function. Thus, we can conclude that our logic gates possess a strong robustness against the disorder of the air-hole missing.

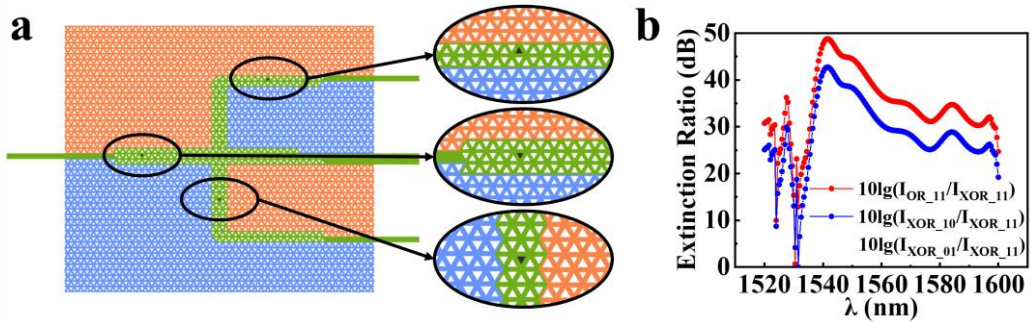

Fig. S9. (a) The schematic diagram of the XOR and OR gates with the disorder of air hole missing. (b) The simulation extinction ratio spectra between logic 1 (XOR_01, XOR_01, and OR_11) and logic 0 (XOR_11).

## S6. The experimental set-up and the measurement method for the logic device.

The experimental set-up is shown in Fig. S10. Our logic chip has four input ports, named as Input A, Input B, $BL_1$, and $BL_2$. The continuous wave laser is used to pump the single mode fiber (SMF) at the communication band (1480nm-1630nm). A quarter-wave plate (QWP) and a half wave plate (HWP) are used to make sure the linear polarization of the pumping laser. Before entering the chip, the incident wave is split into four paths with different intensities in free spaces, where an HWP combined with a polarization beam splitter (PBS) could split the laser beam into two paths with the controlled ratio of intensities. In this case, three PBSs can split the laser beam into four paths (marked as ①, ②, ③, and ④), which are coupled into four SMFs by couplers (black semicircles in Fig. S10). Then, four input lights are guided by SMFs and coupled to the logic chip through the fiber array and 1D gratings.

In experiments, we find that phases of laser beams in four independent SMFs are unstable. It is noted that many factors could affect the phase, including the temperature, the eigenvibration of the fibers, and so on. We find that the phase difference for any pair of SMFs has changed from 0 to $2\pi$ after 20-30 seconds, which is enough to finish the measurement of XOR and OR gates with only two incident signals. But for the cascaded logic gates, more input lights are needed and the stabilized phase differences are extremely hard to be realized. Thus, an alternative method should be used to test the logic chip.

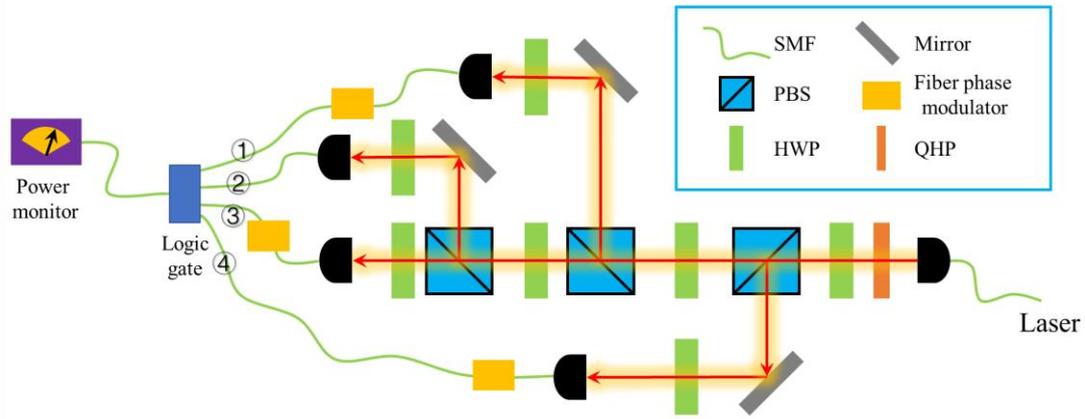

Fig. S10. Experimental set-up of measurement for the logic gate chips.

We add three fiber phase modulators to periodically change the phases of the light in fibers, where the modulation frequencies of three phase modulators are selected from 1Hz to 2kHz, and each modulation frequency is much larger or smaller than others. Moreover, these three frequencies are much larger than that of the spontaneous phase-changed in fibers. In this case, we can think that the phases of signals in fibers are nearly stable in a short interval of time (about 1s). Next, we illustrate the measurement principle in detail [S1].

Firstly, we express the output field from XOR and OR ports as:

$$E_1 = 0.5\sqrt{2}a_A e^{i\pi/2} + 0.5\sqrt{2}a_B e^{i\phi_1(t)}, \tag{S1}$$

$$I_1 = |E_1|^2 = 0.5a_A^2 + 0.5a_B^2 + a_A a_B \sin(\phi_1(t)), \tag{S2}$$

where $E_1$ ($I_1$) is the output field (intensity) at the XOR or OR port. $a_A$ and $a_B$ are amplitudes of incident waves at Input A and B, where we have $a_A = a_B$. As mentioned in the main text, an additional phase (π/2) should be added to the incident signal coming from Input A. $\phi_1(t)$ represents the modulation phase of the signal injected into the Input B. It is noted that $\phi_1(t)$ can be taken as a periodic function along with the time $t$ (like sin, square wave, or sawtooth function). The frequency of $\phi_1(t)$ is set as 10Hz. The difference between the maximum and minimum of $\phi_1(t)$ should equal to $2\pi$, so that the maximum value ($0.5a_A^2 + 0.5a_B^2 + a_A a_B$ corresponds to the output signal of OR_11) and minimum value ($0.5a_A^2 + 0.5a_B^2 + a_A a_B$ corresponds to the output signal of XOR_11) can be obtained when the modulated phase equals to $\phi_1(t) = \pi/2$ and $\phi_1(t) = 3\pi/2$, respectively. In the experiment, a high-speed optical power monitor is used to record the curve of the output intensity, and the measured maximum and minimum values of output signals correspond to the logic state of OR_11 and XOR_11. Additionally, as for other logic states (XOR_10, XOR_01, OR_10, OR_01), only one input signal should be used, making the influence of the spontaneous phase fluctuation could be ignored. Hence, these logic states could be easily detected.

And then, we consider cascaded gates (XNOR, NAND, and NOR), where three input ports (A, B, and BL$_1$) should be used. Except for the phase modulation added on the signal injected into Input B (the same to XOR and OR gates), another phase modulation on the signal injected into the BL$_1$ should also be applied. In this case, the output field ($E_2$) and intensity ($I_2$) from the middle channel could be expressed as:

$$E_2 = 0.5\sqrt{2}E_1 + 0.5\sqrt{2}a_{BL_1}e^{i\phi_2(t)}, \tag{S3}$$

$$\begin{aligned}I_2 &= |E_2|^2 \\ &= 0.25a_A^2 + 0.25a_B^2 + 0.5a_{BL_1}^2 + 0.25a_A a_B \sin(\phi_1(t)) \\ &\quad + 0.25\sqrt{2}a_A a_{BL_1}\sin(\phi_2(t)) + 0.25\sqrt{2}a_B a_{BL_1}\cos(\phi_2(t) - \phi_1(t)),\end{aligned} \tag{S4}$$

where $a_{BL_1}$ is the amplitude of the bias light at BL$_1$. $\phi_2(t)$ represents the modulation phase of the signal injected into BL$_1$. Especially, the modulation frequency of $\phi_2(t)$ is set as 1kHz, which is 100 times larger than that of $\phi_1(t)$.

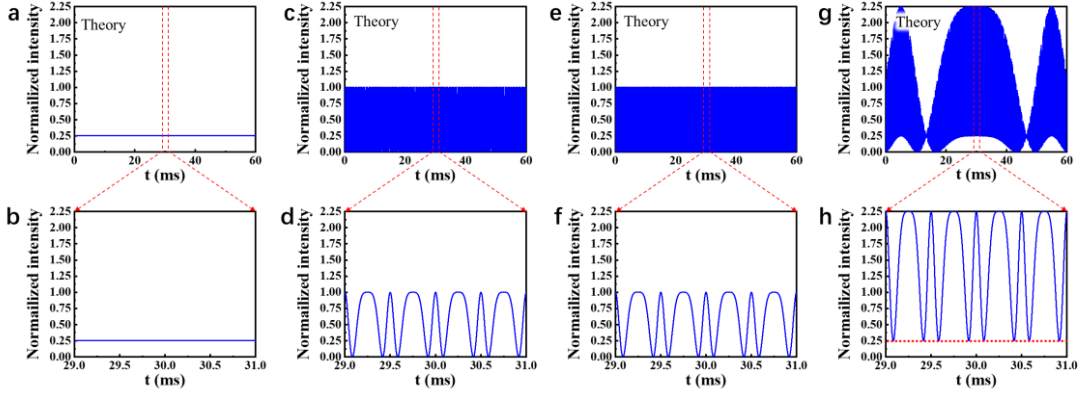

Fig. S11. Theoretical results of the measurement method for the XNOR gate. (a)-(b) Input A and B are both turned-off. (c)-(d) Input A is turned-on but Input B is turned-off. (e)-(f) Input B is turned-on but Input A is turned-off. (g)-(h) Input A and B are both turned-on. (b), (d), (f), and (h) are enlarged views of key parts for (a), (c), (e), and (g), respectively.

We firstly consider the XNOR gate. When Inputs A and B are both turned-off and only the BL$_1$ is turned-on, the output intensity is a constant, i.e. $I_2 = 0.5a_{BL_1}^2$ as shown in Figs. S11(a) and S11(b). When one of Inputs A and B is turned-on, the output intensity is:

$$I_2 = 0.25a_A^2 + 0.5a_{BL_1}^2 + 0.25\sqrt{2}a_A a_{BL_1} \sin(\phi_2(t)),$$

$$\text{or } I_2 = 0.25a_B^2 + 0.5a_{BL_1}^2 + 0.25\sqrt{2}a_B a_{BL_1} \sin(\phi_2(t) - \phi_1(t)),$$

as shown in Figs. S11(c)-S11(f). To realize the logic function, the destructive interference is needed in these two cases. So, we take the minimum of the intensity as the logic signal XNOR_01 (XNOR_10). It is worthy to note that this minimum of the XNOR gate is zero theoretically.

When Inputs A and B are both turned on, the output intensity is expressed as Eq. (S4). In this case, the signal from Input A constructively interferes with that from Input B, and then, the interference signal of Inputs A and B destructively interferes with the signal from BL$_1$. Finally, the output signal arrives at the XNOR port. It is known that the maximum of the intensity curve can be taken when constructive interference appears among these three signals. When the phase of $\phi_2(t)$ changes π, the phase of $\phi_1(t)$ remains unchanged, owing to the much larger modulated frequency of $\phi_2(t)$. That is to say, the expected signal of XNOR_11 can be taken as the local minimum close to the maximum of the curve. As shown in Figs. S11(g) and S11(h), the maximum of the curve is taken at *t*=30ms. The logic signal intensity for XNOR_11 is obtained at t=29.9ms, and its numerical value is 0.25, which is equal to that of XNOR_00 in Fig. S11(a).

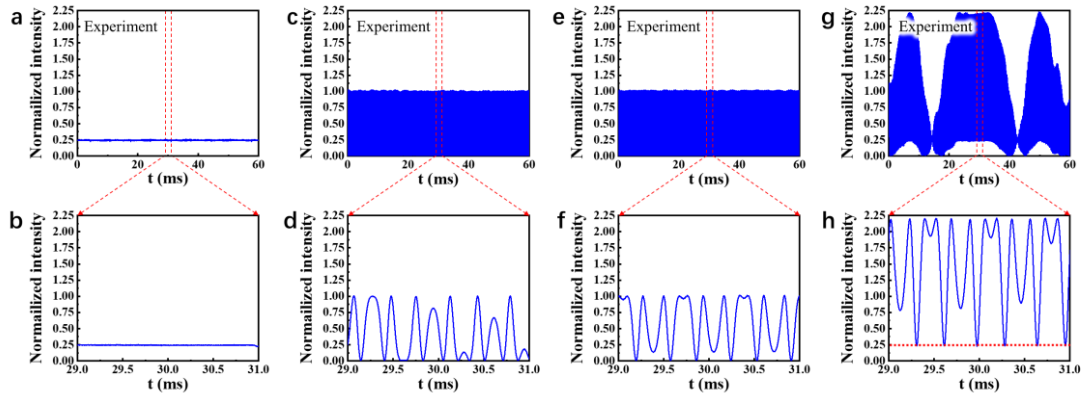

Fig. S12. Experimental results of the measurement method for the XNOR gate. (a)-(b) Input A and B are both turned-off. (c)-(d) Input A is turned-on but Input B is turned-off. (e)-(f) Input B is turned-on but Input A is turned-off. (g)-(h) Input A and B are both turned-on. (b), (d), (f), and (h) are enlarged views of key parts for (a), (c), (e), and (g), respectively.

In Figs. S12(a), S12(c), S12(e), and S12(g), we show experimental intensity curves corresponding to four logic cases as described above. And Figs. S12(b), S12(d), S12(f), and S12(h) display the enlarged views of curves. It can be seen measured results are in good agreement with theoretical results. We take output signals for the XNOR gate at various wavelengths (from 1520nm to 1605nm), and then we calculate and plot ERs in Fig. 3 of the main text.

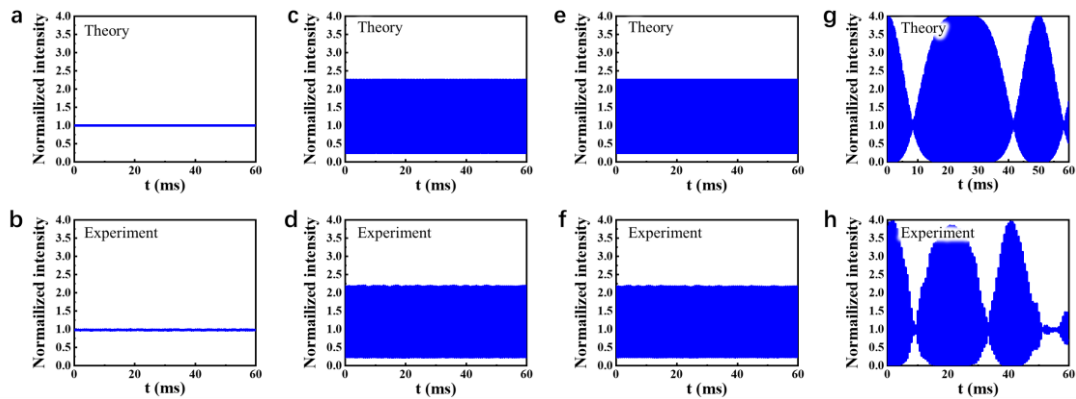

Fig. S13. Theoretical and experimental results of the measurement method for the NAND gate. (a)-(b) Input A and B are both turned-off. (c)-(d) Input A is turned-on but Input B is turned-off. (e)-(f) Input B is turned-on but Input A is turned-off. (g)-(h) Input A and B are both turned-on. (a), (c), (e), and (g) are theoretical results, (b), (d), (f), and (h) are experimental results, respectively.

As for the NAND gate, Eqs. S3 and S4 can also be applied except for the amplitude of the bias light at $BL_1$ being changed to $\sqrt{2}$. We can get output intensities of logic signals in the same way. Theoretical and experimental intensity curves are shown in Fig. S13. Four logic states correspond to Figs. S13(a) and S13(b), Figs. S13(c) and S13(d), Figs. S13(e) and S13(f), Figs. S13(g) and S13(h), respectively. It is worthy to note that the NAND_11 case shows the output intensity is zero, which corresponds to the logic function of the NAND gate.

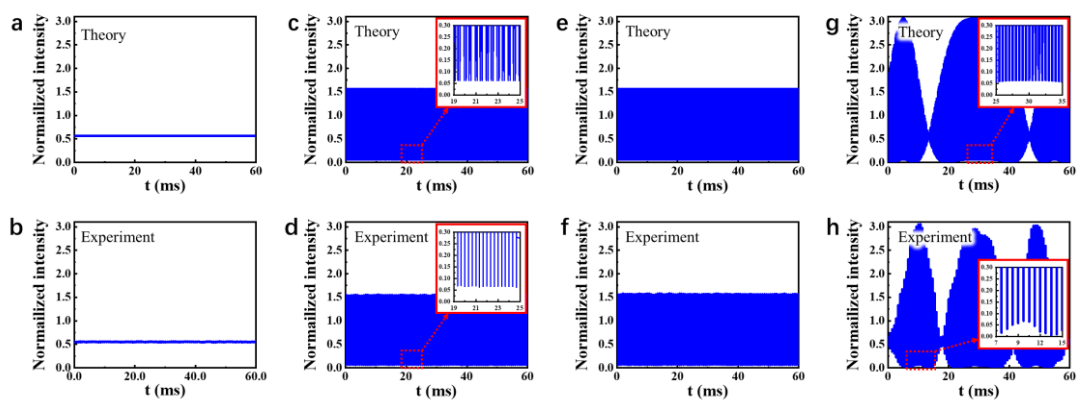

Fig. S14. Theoretical and experimental results of the measurement method for the NOR gate. (a)-(b) Input A and B are both turned-off. (c)-(d) Input A is turned-on but Input B is turned-off. (e)-(f) Input B is turned-on but Input A is turned-off. (g)-(h) Input A and B are both turned-on. (a), (c), (e), and (g) are theoretical results, (b), (d), (f), and (h) are experimental results, respectively.

In Fig. S14, we plot theoretical and experimental results for the NOR gate. The measurement method in such a case is identical with the cases for XNOR and NAND gates, except for the amplitude of the bias light at $BL_1$ being changed to $0.75\sqrt{2}$. The output intensity being 0.5625 (logic signal 1) can be obtained when Inputs A and B are both turned-off, as shown in Figs. S14(a) and S14(b). In the other three cases (Figs. S14(c)-S14(h)), the logic output intensity can be calculated and measured as 0.0625 (1/9 of 0.5625). The experiment measurement corresponds well to the theoretical results.

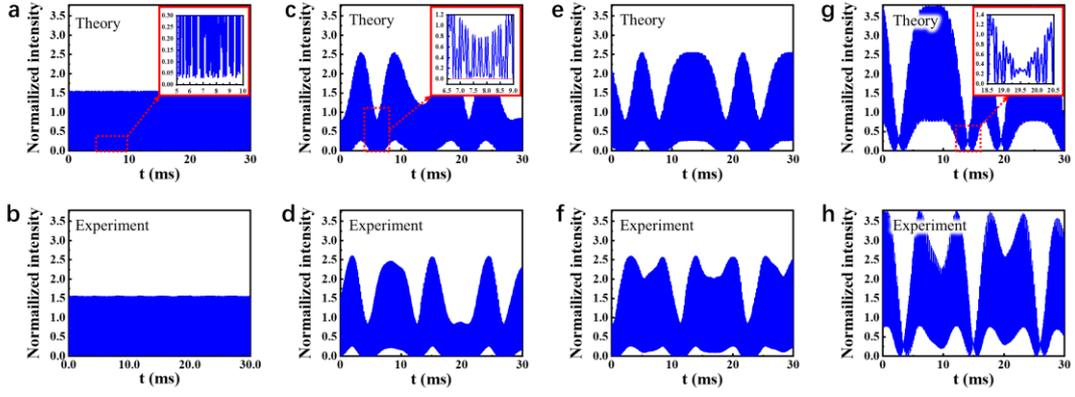

Fig. S15. Theoretical and experimental results of the measurement method for the AND gate. (a)-(b) Input A and B are both turned-off. (c)-(d) Input A is turned-on but Input B is turned-off. (e)-(f) Input B is turned-on but Input A is turned-off. (g)-(h) Input A and B are both turned-on. (a), (c), (e), and (g) are theoretical results, (b), (d), (f), and (h) are experimental results, respectively.

In the last case for the AND gate, additional formulas are needed as follows:

$$E_3 = 0.5\sqrt{2}E_2 + 0.5\sqrt{2}a_{BL_2}e^{i\phi_3(t)}, \tag{S5}$$

$$I_3 = |E_3|^2 = 0.125a_A^2 + 0.125a_B^2 + 0.25a_{BL_1}^2 + 0.5a_{BL_2}^2$$

$$+ 0.125a_A a_B \sin(\phi_1(t)) + 0.125\sqrt{2}a_A a_{BL_1}\sin(\phi_2(t)) + 0.25a_A a_{BL_2}\sin(\phi_3(t))$$

$$+ 0.125\sqrt{2}a_B a_{BL_1}\cos(\phi_2(t) - \phi_1(t)) + 0.25a_B a_{BL_2}\cos(\phi_3(t) - \phi_1(t))$$

$$+ 0.25\sqrt{2}a_{BL_1}a_{BL_2}\cos(\phi_3(t) - \phi_2(t)), \tag{S6}$$

where $E_3$ ($I_3$) is the field (intensity) of output signals at the AND port, $a_{BL_2}$ is the amplitude of the bias light at $BL_2$. $\phi_3(t)$ represents the modulation phase of the signal injected into $BL_2$. We set the frequency of $\phi_3(t)$ as 2kHz, and that of $\phi_1(t)$ ($\phi_2(t)$) is changed to 1Hz (30Hz).

When Inputs A and B are both turned-off, the $BL_1$ and $BL_2$ are turned-on. The output intensity $I_3$ can be expressed as: $0.25a_{BL_1}^2 + 0.5a_{BL_2}^2 + 0.25\sqrt{2}a_{BL_1}a_{BL_2}\cos(\phi_3(t) - \phi_2(t))$. At the AND port, the output waves from $BL_1$ and $BL_2$ need to destructively interfere with each other for realizing the logic function of the AND gate. Thus, we take the signal intensity from the minimum of the curve of Fig. S15(a). Here, the normalized output intensity for the logic signal 0 is 0.03.

When one of Inputs A and B is turned-on and another is turned-off, the output intensity is:

$$I_3 = 0.125a_A^2 + 0.25a_{BL_1}^2 + 0.5a_{BL_2}^2 + 0.125\sqrt{2}a_A a_{BL_1}\sin(\phi_2(t)) + 0.25a_A a_{BL_2}\sin(\phi_3(t)), \text{ or}$$

$$I_3 = 0.125a_B^2 + 0.25a_{BL_1}^2 + 0.5a_{BL_2}^2 + 0.125\sqrt{2}a_B a_{BL_1}\cos(\phi_2(t)-\phi_1(t)) + 0.25a_B a_{BL_2}\cos(\phi_3(t)-\phi_1(t)).$$

In this case, the signal from Input A(B) destructively interferes with that from $BL_1$, and then, the interference signal of Input A and $BL_1$ destructively interferes with the signal from $BL_2$. Finally, the output signal arrives at the AND port. Thus, we can take the signal intensity of logic states (AND_01 and AND_10) as the local value of the bottom envelope of the curve close to the minimum of the top envelope. In Figs. S15(c) and S15(e), the logic signal 0 can be obtained, whose value is 0.03.

When both Inputs A and B are turned-on (AND_11), the output intensity is expressed as Eq. (S6). In this case, the signal from Input A constructively interferes with that from Input B, and then, the interference signal of Inputs A and B destructively interferes with the signal from $BL_1$. Next, the interference signal of Inputs A, B, and $BL_1$ destructively interferes with the signal from $BL_2$. Finally, the output signal arrives at the AND port. Thus, the output intensity of the logic signal (AND_11) can be also taken as the local value of the bottom envelope of the curve close to the minimum of the top envelope (the cross point of the top and bottom envelopes), as shown in Fig. S15(g). The signal intensity is about 0.27. The intensity ratio of logic signals 1 and 0 is 9:1 (ER=9.54dB). Such experimental results can be seen in Figs. S15(b), S15(d), S15(e), and S15(h). They are in a good agreement with theoretical results.

In all, according to the above measurement method, we can obtain all logic signals from logic gates in the communication band. All measured results are shown in Figs. 2-4 of the main text.

**S7. The scheme of the optical phase lock loops (OPLLs).**

Here, we propose a scheme to stabilize the phase difference between logic input signals and the bias lights. The optical phase lock loops (OPLLs) [S2-S4] are such a way to offset the phase fluctuations. As shown in Fig. S16, we use three fiber tunable couplers to split the laser into four paths. The strengths of the input light in every fiber can be flexibly adjusted, so the amplitude ratio of input signals for all logic gates is satisfied.

The phase of logic input A is chosen as the phase standard. The phases of the other three input signals are adjusted by a feedback method in the OPLLs. On the SOI chip, input signal A is split into four paths, and input signals B, $BL_1$, $BL_2$ are split into two paths. They interfere in pairs at Y branches, and then the interference signals are exported and injected into the OPLLs. The interference optical output signals are converted to electrical signals. After the signal processing in the OPLLs, the corresponding output electrical signals are injected into the fiber phase modulators. So, the phases of inputs B, $BL_1$, $BL_2$ can be fixed. By such a feedback adjustment, the phase differences of input signals can be stable. Thus, all logic gates we designed can be used for a long time.

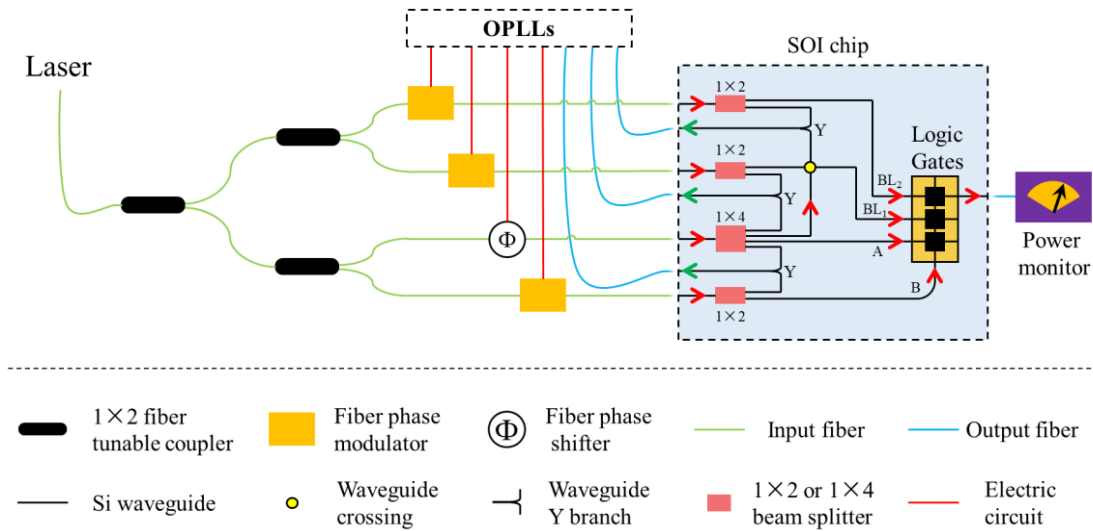

**Fig. S16.** The schematic diagram of the logic gate measurement by using the optical phase lock loops (OPLLs).